\documentclass{ifacconf}

\usepackage{graphicx}      
\usepackage{natbib}        
\usepackage{todonotes}     
\presetkeys{todonotes}{inline,fancyline}{} 
\usepackage{amssymb,amsmath,amsbsy} 
\usepackage{tabu} 
\usepackage{url} 
\usepackage{tikz}
\usetikzlibrary{positioning}
\usetikzlibrary{arrows}
\usetikzlibrary{backgrounds}
\usetikzlibrary{calc}
\usetikzlibrary{decorations.pathmorphing}
\usetikzlibrary{shapes}
\usetikzlibrary{automata}
\def\a{15} 
\usepackage{pgfplots} 
\pgfplotsset{compat=1.17}
\usepgfplotslibrary{external} 
\usepackage{uppaal} 

\makeatletter
\renewcommand{\todo}[2][]{\tikzexternaldisable\@todo[#1]{#2}\tikzexternalenable}
\makeatother

\DeclareMathOperator*{\argmin}{arg\,min} 

\newcommand{\hl}[1]{\textcolor{black}{#1}}
\newcommand{\hll}[1]{\textcolor{black}{#1}}
\newcommand{\hlr}[1]{\textcolor{black}{#1}}

\begin{document}
\begin{frontmatter}

\title{Learning Safe and Optimal Control Strategies for Storm Water Detention Ponds\thanksref{footnoteinfo}} 

\thanks[footnoteinfo]{We thank the support from Villum Synergy project CLAIRE and ERC Advanced Grant LASSO.}

\author[CS]{Martijn A. Goorden}
\author[CS]{Kim G. Larsen}
\author[CE]{Jesper E. Nielsen}
\author[CS]{Thomas D. Nielsen}
\author[CE]{Michael R. Rasmussen}
\author[CS]{Ji\v{r}\'{i} Srba}

\address[CS]{Department of Computer Science, Aalborg University, 
   Aalborg, Denmark (e-mail: \{mgoorden,kgl,tdn,srba\}@cs.aau.dk).}
\address[CE]{Department of Built Environment, Aalborg University,
   Aalborg, Denmark, (e-mail: \{jen,mrr\}@build.aau.dk)}

\begin{abstract}                
Storm water detention ponds are used to manage the discharge of rainfall runoff from urban areas to nearby streams. Their purpose is to reduce the hydraulic impact and sediment loads of the receiving waters. Detention ponds are currently designed based on static controls: the output flow of a pond is capped at a fixed value. This is not optimal with respect to the current infrastructure capacity and for some detention ponds it might even violate current regulations set by the European Water Framework Directive. We apply formal methods to synthesize (i.e., derive automatically) a safe and optimal active controller. We model the storm water detention pond, including the urban catchment area and the rain forecasts, as a hybrid Markov decision process. Subsequently, we use the tool \stratego to synthesize a control strategy minimizing the cost related to pollution (optimality) while guaranteeing no emergency overflow of the detention pond (safety). Simulation results for an existing pond show that \stratego can learn \hll{optimal} strategies that prevent emergency overflows, where the current static control is not always able to prevent it. \hll{At the same time, our approach can improve sedimentation during low rain periods.}
\end{abstract}


\end{frontmatter}

\section{Introduction}\label{sect:introduction}
Urbanization poses two major risks related to storm water runoff management: flooding of the urban area, and  \hll{environmental impact on receiving waters from hydraulic loads and pollution}. \hll{The roads, roofs and other man-made surfaces of urban areas collect the rainwater and generate runoff, which needs to be transported away from the city to receiving waters to avoid urban flooding. The urban runoff carries particulates and xenobiotics from depositions on the urban surfaces~\citep{hvitved-jacobsen_urban_2010}.} Storm water detention ponds are the most commonly used storm water management tool for avoiding or reducing the impacts of storm water runoff~\citep{tixier_ecological_2011}. Negative impacts of storm water runoff include unnatural disturbances, stream bed erosion, and downstream flooding~\citep{walsh_urban_2005}.

With the growth of urban areas as well as the change in climate and its related rain events, water utility companies need to constantly construct, maintain, and upgrade detention ponds to ensure efficient and safe storm water discharge. For example, it is estimated by the Danish Water and Wastewater Association that the cost of updating the Danish storm water systems will be between 0.6 and 1.3 billion euros~\citep{danva_report_2018}. \hll{Mitigating the environmental impact has increasingly changed the discharge permits over the last 15 years, so even recently constructed storm water detention ponds do not comply with present recommendations, see~\cite{jensen_eu_2020}.}

The requirements for the design of storm water detention ponds are, in general, based on the maximum allowed discharge flow rate into the nearby stream or river, the probability of emergency overflow, \hll{and the concentration of pollutants in the discharged water}~\citep{mobley_design_2014}. These regulations are derived from the European Water Framework Directive~\citep{eu_2000} and discharge permits issued by the local authorities.

Currently, the most common discharge strategies involve static flows (when there is storm water in the detention pond) into the stream without taking into account the actual capacity of the receiving stream. These strategies do not necessarily comply with the regulations set by the European Water Framework Directive, as pointed out by~\cite{schutze_real_2004}.

Existing work on the design of active control strategies for urban water systems primarily focus on sewer systems and wastewater treatment plants, which do include detention ponds as a subsystem, see for example~\cite{schutze_real_2004,haverkort_evaluating_2010,hoppe_real-time_2011,sun_real-time_2017}. Controlled discharge from detention ponds has also been studied. In~\cite{muschalla_ecohydraulic-driven_2014}, a real-time controller is designed improving the efficiency of particle removal. Further improvements are presented in~\cite{gaborit_improving_2013,gaborit_exploring_2016}, where off-line strategies take weather forecasts into account. However, these works focus primarily on particle removal efficiency as the objective, \hll{as most of the pollution is bound strongly to organic and inorganic particles}. Furthermore, the underlying rule-based control strategies are manually derived and the discharge output can only be changed infrequently, like once a day.

In the field of cyber-physical systems, there is considerable research in deploying formal methods for verifying or even synthesizing controllers for stochastic hybrid systems. For example, the tool \stratego~\citep{david_uppaal_2015} is able to synthesize safe and near-optimal controllers by combining model checking and reinforcement learning. Case studies using \stratego include battery aware scheduling problems~\citep{wognsen_score_2015}, adaptive cruise control~\citep{larsen_safe_2015}, and floor heating~\citep{larsen_online_2016}.

In this paper, we deploy formal methods to automatically synthesize active control strategies for storm water detention ponds, where we optimize for particle sedimentation while also guaranteeing safety with respect to emergency overflows. As basis for the control synthesis, we model the storm water detention pond as a hybrid Markov decision process utilizing a combination of differential equations and (stochastic) timed automata. The synthesized controller is able to periodically change the discharge flow, thus allowing a more rapid and precise response to uncertain weather events. \hll{Simulation experiments of a real-world detention pond show that} the synthesized strategies are better at utilizing the pond's capacity compared to the current static control \hll{while also honoring the stated safety requirements}.

\section{Storm water detention ponds}\label{sect:systemdescription}
\begin{figure}
	\includegraphics[width=\linewidth]{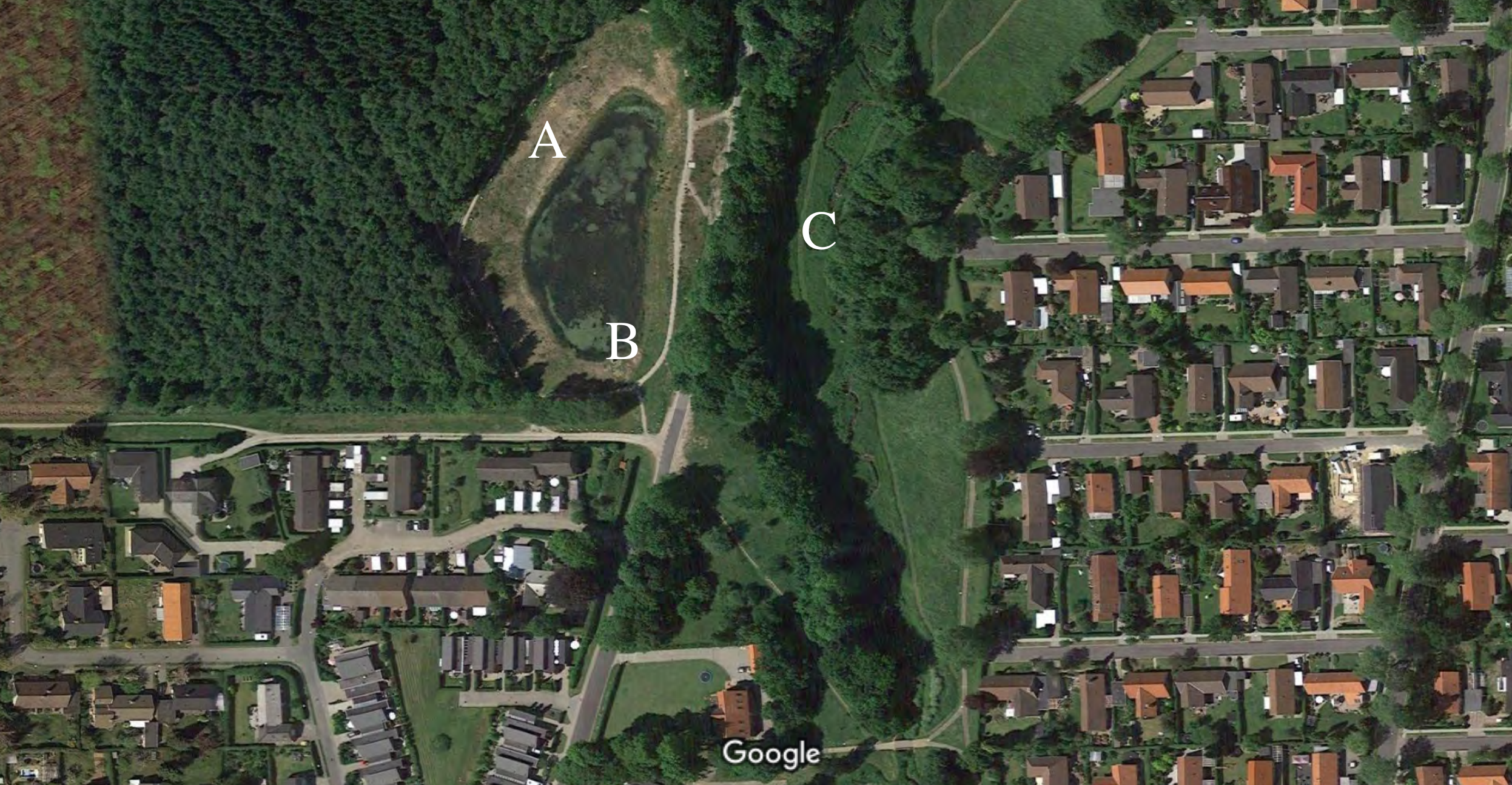}
	\caption{Satellite image of the \hll{Vilhelmsborg Skov pond south of Aarhus, Denmark}. A is the detention pond, where the outlet is located at B. Storm water is discharged in the stream labeled with C, which runs from the south to the north. Image from Google Maps.}\label{fig:aaupond}
\end{figure}

Storm water detention ponds collect storm water from urban areas, like streets, roofs, parking lots, and recreational parks. When it is raining or snow is melting, two main risks arise. First, the runoff flow in urban areas can exceed the capacity of nearby streams. Second, urban area particles collected by the storm water can pollute the ecosystem of the stream and downstream waterbodies. Storm water detention ponds aim to reduce or avoid the impact of both risks.

An example of such a pond is shown in Figure~\ref{fig:aaupond}. The satellite image shows \hll{the Vilhelmsborg Skov pond south of Aarhus, Denmark}. Labeled by A is the storm water detention pond, \hll{partially filled with water}. It collects the water from from \hll{the neighborhoods south of it} through the storm water sewer system next to the roads. The pond's outlet is indicated by B, which connects to the stream labeled with C. This stream runs from south to north and discharges the water from \hll{the neighborhood}.

Storm water detention ponds can be characterized by being either a wet or a dry detention pond. A wet detention pond always has a minimal amount of water in it, while a dry one can empty completely (hence the names). In our case study, we focus on a wet detention pond.

Currently, storm water detention ponds are designed with static outlet flow regulator creating a capped outlet flow into the stream. The capacity of the stream (reflected in the issued permits) dictates the maximum outlet flow of the pond. This capped outlet flow determines, together with other criteria like the urban catchment size and the emergency overflow risk, the size of the pond. 

Recent research has focused on the design of energy efficient dynamic outlet flow regulator\footnote{\hll{See project webpage at \url{https://www.danva.dk/viden/vudp/projektuddelinger/relevand/} (in Danish).}}. Having these flow regulators allows utility companies to incorporate active (or real-time) control into their design of the storm water detention ponds. This has the potential to enable more efficient detention pond designs and reduce the negative effect of the two aforementioned main risks.

\begin{figure}
	\centering
	\begin{tikzpicture}[scale=.6, transform shape]		
		\node[coordinate] (lb) {};
		\node[coordinate] (rb) [right = 4cm of lb] {};
		\node[coordinate] (lt) [above left = 3cm and 1.5cm of lb] {};
		\node[coordinate] (rt) [above right = 3cm and 1.5cm of rb] {};
		
		\node[] (P1) [above = .1 cm of lb] {};
		\node[] (P2) [above = .3cm of P1] {};
		\node[coordinate] (irb) at (intersection of  lb--lt and P1.east--P1.west) {};
		\node[coordinate] (irt) at (intersection of  lb--lt and P2.east--P2.west) {};
		\node[coordinate] (ilb) [left = 1.5cm of irb] {};
		\node[coordinate] (ilt) at (ilb |- irt) {};
		
		\draw
		(ilb) -- (irb)
		(ilt) -- (irt);
		
		\begin{scope}[on background layer]
			\fill[blue!20!white]
			(ilb) -- (irb) -- (irt) -- (ilt) -- cycle;	
		\end{scope}	
		
		\path (ilb) -- node[coordinate, pos=.5] (iae) {} (ilt);
		\node[] (iab) [left = .6cm of iae] {\Large$Q_{\mathit{in}}$};
		
		\draw[->, >=stealth'] (iab) -- (iae);
		
		\node[] (Q1) [above = 0.75 cm of rb] {};
		\node[] (Q2) [above = .3cm of Q1] {};
		\node[coordinate] (olb) at (intersection of  rb--rt and Q1.east--Q1.west) {};
		\node[coordinate] (olt) at (intersection of  rb--rt and Q2.east--Q2.west) {};
		\node[coordinate] (orb) [right = 1.5cm of olb] {};
		\node[coordinate] (ort) at (orb |- olt) {};
		
		\draw
		(olb) -- (orb)
		(olt) -- (ort);
		
		\begin{scope}[on background layer]
			\fill[blue!20!white]
			(olb) -- (orb) -- (ort) -- (olt) -- cycle;
		\end{scope}
		
		\path (orb) -- node[coordinate, pos=.5] (oae) {} (ort);
		\node[] (oab) [right = .3cm of oae] {\Large$Q_{\mathit{out}}$};
		
		\draw[<-, >=stealth'] (oab) -- (oae);
		
		\draw
		(lt) -- (irt)
		(irb) -- (lb)
		(lb) -- (rb) 
		(rb) -- (olb)
		(olt) -- (rt);
		
		\draw[dashed, color=gray] (olb) -- (intersection of lb--lt and olb--orb);
		
		\draw[dashed, color=gray] (lt) -- (rt);
		
		\node[] (wl) [above = 2.3cm of lb] {};
		\node[coordinate] (lwl) at (intersection of lb--lt and wl.east--wl.west) {};
		\node[coordinate] (rwl) at (intersection of rb--rt and wl.east--wl.west) {};
		\draw[decorate, decoration={coil,aspect=0,segment length=0.7cm}] (rwl) -- (lwl);
		
		\begin{scope}[on background layer]
			\path[fill,blue!20!white]
			decorate[decoration={coil,aspect=0,segment length=0.7cm}] {(rwl) -- (lwl)} 
			[]-- (lwl) -- (lb) -- (rb) -- cycle;
		\end{scope}
		
		\node[] (mb) [right=0.5cm of lb] {};
		\path[<->, >=stealth',shorten >=3pt, shorten <=1pt] (intersection of mb.north--mb.south and olb--orb) edge node[right] {\Large$w$} (intersection of mb.north--mb.south and wl.east--wl.west);
		
		\node[] (mb2) [left=0.5cm of rb] {};
		\path[<->, >=stealth',shorten >=1pt, shorten <=1pt] (intersection of mb2.north--mb2.south and olb--orb) edge node[right] {\Large$W$} (intersection of mb2.north--mb2.south and lt--rt);
		
		\node[] (vl) [right = 1cm of olb] {};
		\node[coordinate] (vh) at (intersection of olt--ort and vl.south--vl.north) {};
		\draw[line width=2pt] (vl.center) -- (vh);
		\node[] (v) [above = .1cm of vh] {Valve};
		
		%
		%
		
		\node[draw, minimum height=3cm, minimum width=3cm] (ua) [above left = -0.9cm and 1cm of iab] {\Large Urban area};
		\node[] (rain) [above = .8cm of ua] {\Large Rain};
		
		\path[-] (ua.east |- iab) edge (iab);
		\path[->, >=stealth', shorten >=1pt] (rain) edge (ua.north);		
	\end{tikzpicture}
	\caption{Overview of the storm water detention pond}\label{fig:systemoverview}
\end{figure}
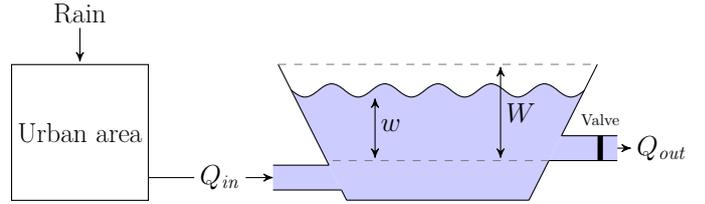

Figure~\ref{fig:systemoverview} shows a schematic overview of the wet storm water detention pond. Rain water falls into an urban area, like a neighborhood, university campus area, or a highway, and is transported to a nearby pond. Rain water enters the pond through inlet $Q_{\mathit{in}}$ and exits it through outlet $Q_{\mathit{out}}$ into a nearby stream or river. A valve in the outlet limits the outflow. Due to the positioning of the outlet pipe, there is a permanent water level in the pond (indicated with the lower horizontal dashed line in the figure). The variation of the water level above this permanent water level is indicated by $w$. \hll{When the maximum water level, indicated by $W$, is reached, the pond will overflow.}

We neglect the rain falling directly on the pond surface, the water evaporating into the atmosphere, the water infiltration into the pond from the bottom and the sides, and the water leaking through the bottom and sides. It is shown in~\cite{thomsen_simplified_2019} that these water flows are negligible compared to the storm water flow.

\section{Preliminaries}\label{sect:preliminaries}

We use the mathematical modeling framework of hybrid Markov decision process (HMDP), \hll{adapted from~\cite{ashok_sos_2019,larsen_online_2016}}.

\begin{defn}
	A \emph{hybrid Markov decision process} (HMDP) $\mathcal{M}$ is a tuple $(C,U,X,F,\delta)$ where:
	\begin{itemize}
		\item the controller $C$ is a finite set of (controllable) modes $C = \{c_1, \ldots, c_k\}$,
		\item the uncontrollable environment $U$ is a finite set of (uncontrollable) modes $U=\{u_1,\ldots,u_l\}$,
		\item $X = \{x_1, \ldots, x_n\}$ is a finite set of continuous (real-valued) variables,
		\item for each $c\in C$ and $u\in U$, the flow function $F_{c,u}:\mathbb{R}_{>0}\times \mathbb{R}^X \rightarrow \mathbb{R}^X$ describes the evolution of the continuous variables over time in the combined mode $(c,u)$, and
		\item $\delta$ is a family of \hll{density} functions $\delta_{\gamma}:\hll{\mathbb{R}_{\geq 0}\times} U\rightarrow [0,1]$, where $\gamma = (c,u,\boldsymbol{x})$ is a global configuration with $\boldsymbol{x} : X \rightarrow \mathbb{R}$ being a valuation. More precisely, $\delta_{\gamma}(\hll{\tau},u')$ is the probability that in the global configuration $(c,u,\boldsymbol{x})$ the uncontrollable mode $u$ will change to mode $u'$ \hll{after a delay $\tau$}\footnote{\hll{Note that $\Sigma_{u'}\int_{\tau} \delta_{\gamma}(\tau,u') \mathrm{d}\tau = 1$.}}.
	\end{itemize}
\end{defn}

This notion of an HMDP describes an \hlr{uncountable and infinite} state Markov Decision Process, see~\cite{puterman_markov_1994}, where the controller mode switches periodically \hll{with interval $P\in\mathbb{R}_{\geq 0}$} and the uncontrollable environment mode switches probabilistically according to $\delta$. In the rest of the paper, we denote by $\mathbb{C}$ the set of global configurations $C\times U\times (X \rightarrow \mathbb{R})$ of an HMDP.

The evolution of an HMDP over time is defined as follows. Let $\gamma=(c,u,\boldsymbol{x})$ be the current configuration, $\gamma'=(c',u',\boldsymbol{x}')$ the next configuration, and $\tau$ a time delay. We write $\gamma \xrightarrow{\tau}\gamma'$ when $c'=c$, $u'=u$, and $\boldsymbol{x}' = F_{c,u}(\tau, \boldsymbol{x})$. \hll{We write $\gamma \xrightarrow{\tau}_{u}\gamma'$ in case $c' = c$, $\boldsymbol{x}' = F_{c,u}(\tau, \boldsymbol{x})$ and $\delta_{\gamma}(\tau, u') > 0$.}

A \emph{run} of an HMDP is an interleaved sequence $\pi\in \mathbb{C}\times(\mathbb{R}_{\geq 0}\times \mathbb{C})^*$ of configurations and time-delays, starting with initial configuration $\gamma_0$:
\begin{equation*}
	\pi = \gamma_0 :: \hll{\tau_1}::\gamma_1::\hll{\tau_2}::\gamma_2::\hll{\tau_3}\cdots
\end{equation*}
\hl{where} \hll{$\gamma_i=(c_i, u_i, \boldsymbol{x}_i)$, for all $n$ there exist $i$ such that $\Sigma_{j\leq i} \tau_j = n \cdot P$, and for all $i$ either}
\begin{enumerate}
	\item the environment changes to a new mode, i.e., \hll{$\gamma_i \xrightarrow{\tau_{i+1}}_{u}\gamma_{i+1}$}, or
	\item the controller changes to any possible new mode \hll{when it reaches the end of a period}, i.e., \hll{$\Sigma_{j\leq i+1}\tau_j$ is a multiple of $P$ and $\gamma_i\xrightarrow{\tau_{i+1}} (c_i, u_i, \boldsymbol{x}_{i+1})$ with $c_{i+1}\in C$ and $u_{i+1} = u_i$}.
\end{enumerate}

\hl{The control problem of a storm water detention pond can be described as an HMDP as follows. The set of control modes $C$ contains the different pond outlet valve settings that can be chosen. For static control, $C$ becomes a singleton. The rain determines the uncontrollable stochastic input to the system, \hll{with two uncontrollable modes for dry and raining. The density function $\delta$ captures the uncertainty in the duration of the dry and rain intervals, which is independent of $c$ and $\boldsymbol{x}$.} Finally, $X$ contains the state variables, such as the current water level in the pond and the current rain intensity.}

\hl{For the model of the detention pond, the flow function $F_{c,u}$ is expressed as a combination of differential equations and timed automata. A timed automaton~\citep{behrmann_tutorial_2004} is a tuple $A=(\mathit{Loc}, l_0, \mathit{Clk}, E, \mathit{Act}, \mathit{Inv})$, where $\mathit{Loc}$ is a finite set of locations, $l_0\in\mathit{Loc}$ is the initial location, $\mathit{Clk}$ is a finite set of clocks, $E\subseteq \mathit{Loc}\times\mathit{Act}\times\mathcal{B}(\mathit{Clk})\times 2^{\mathit{Clk}} \times\mathit{Loc}$ is a set of edges, $\mathit{Act}$ is a finite set of actions, and $\mathit{Inv} : \mathit{Inv}\mapsto \mathcal{B}(\mathit{Clk})$. Here $\mathcal{B}(\mathit{Clk})$ is the set of all predicates using the clocks. When such a predicate is used on an edge, it is called a guard. Finally, $2^{\mathit{Clk}}$ indicates on edges the set of clocks that are reset to 0.}

\subsubsection{\hll{Example}}

\begin{figure}
	\begin{center}
	\begin{tikzpicture}
		\node[coordinate] (lb) {};
		\node[coordinate] (rb) [right = 4cm of lb] {};
		\node[coordinate] (lt) [above = 2cm of lb] {};
		\node[coordinate] (rt) [above = 2cm of rb] {};
		
		\draw[thick]
		(lt) -- node[coordinate, pos=.4] (lm) {} (lb) -- node[coordinate, pos=.5] (mb) {}  (rb) -- node[coordinate, pos=.6] (rm) {} (rt);
		
		\path (lt) -- node[coordinate, pos=.5] (mt) {} (rt);
		\node (rain) [above = .8cm of mt] {Rain};
		\path[->, >=stealth', thick] (rain) edge (mt);	
		
		\draw (lm) -- (rm);
		\node[] (mb2) [right=1cm of lb] {};
		\path[<->, >=stealth',shorten >=1pt, shorten <=1pt] (intersection of mb2.north--mb2.south and lb--rb) edge node[right] {$V$} (intersection of mb2.north--mb2.south and lm--rm);
		
		\node[coordinate] (mbl) [left = .3cm of mb] {};
		\node[coordinate] (mbr) [right = .3cm of mb] {};
		\node[coordinate] (ol) [below = 1cm of mbl] {};
		\node[coordinate] (or) [below = 1cm of mbr] {};
		\draw[thick] 
		(mbl) -- node[coordinate, pos=.2] (lv) {} (ol)
		(mbr) -- node[coordinate, pos=.2] (rv) {} (or);
		
		\draw[fill]
		(mbl) -- (lv) -- (rv) -- (mbr) -- cycle;
		
		\node[coordinate] (lwm) [above = .4cm of lb] {};
		\node[coordinate] (rwm) [above = .4cm of rb] {};
		\draw[dash pattern=on5pt off3pt] (lwm) -- node[pos=.75, fill=white] (lm) {$V_{\mathit{min}}$}  (rwm);
		
		\node[coordinate] (w) [below left = 0.2cm and 0.2cm of rm] {};
		\node (water) [right = 1cm of w] {Water};
		\path[color=white!70!black, thick] (water) edge (w);
		
		\node (c) [below left = 0.1cm and 0.1cm of mbr] {};
		\node (control) at (intersection of water.north--water.south and c.north west--c.north east) {Controller};
		\path[color=white!70!black, thick] (control) edge (c.north east);	
	\end{tikzpicture}
	\end{center}
	\emph{Rain}
	\begin{center}
	\begin{tikzpicture}[->,>=stealth',shorten >=1pt,opacity=1, auto,node distance=8cm,
		thick,main node/.style={circle,draw,font=\sffamily\Large\bfseries},scale=0.85,transform shape]
		\node[initial,  initial text ={}, initial where=left, main node, label={above:$\texttt{Dry}$}, label={below:$x \leq 12$}, minimum size=10pt] (1) {};
		
		\node[main node,  label={above:$\texttt{Raining}$}, label={below:$x \leq 12$}, minimum size=10pt] (2) [right of=1] {};
		
		\path[every node/.style={font=\sffamily}, dash pattern=on5pt off3pt, bend left=10, align=center]
		(1) edge [bend left] node [above] {$x\geq 6$\\ $x :=0, r \in \mathit{uniform}(5,10)$} 	(2)
		(2) edge [bend left] node [below] {$x\geq 8$ \\  $x := 0, r := 0.0$} (1);		
	\end{tikzpicture}
	\end{center}
	\emph{Controller}
	\begin{center}
	\begin{tikzpicture}[->,>=stealth',shorten >=1pt,opacity=1, auto,node distance=4cm,
		thick,main node/.style={circle,draw,font=\sffamily\Large\bfseries},scale=0.85,transform shape]
		\node[initial,  initial text ={}, initial where=left, main node, label={above:$\texttt{Closed}$}, label={below:$y \leq 1$}, minimum size=10pt] (1) {};
		
		\node[main node,  label={above:$\texttt{Open}$}, label={below:$y \leq 1$}, minimum size=10pt] (2) [right of=1] {};
		
		\path[every node/.style={font=\sffamily}, bend left=10, align=center]
		(1) edge [bend left] node [above] {$y = 1$ \\ $y :=0, o := 8$} 	(2)
		(2) edge [bend left] node [below] {$y = 1$ \\  $y := 0, o := 0$} (1);		
		
		\path[every node/.style={font=\sffamily}, align=center]
		(1) edge [loop left, looseness=25, out=210, in=150] node [left] {$y = 1$ \\  $y := 0$} (1)
		(2) edge [loop right, looseness=25, out=30, in=330] node [right] {$y = 1$ \\ $y :=0$} (2);
	\end{tikzpicture}
	\end{center}
	\emph{Water}
	\begin{equation*}
		\frac{\mathrm{d}V}{\mathrm{d}t} = r - o
	\end{equation*}
	\caption{\hll{A small HMDP example.}}\label{fig:example}
\end{figure}
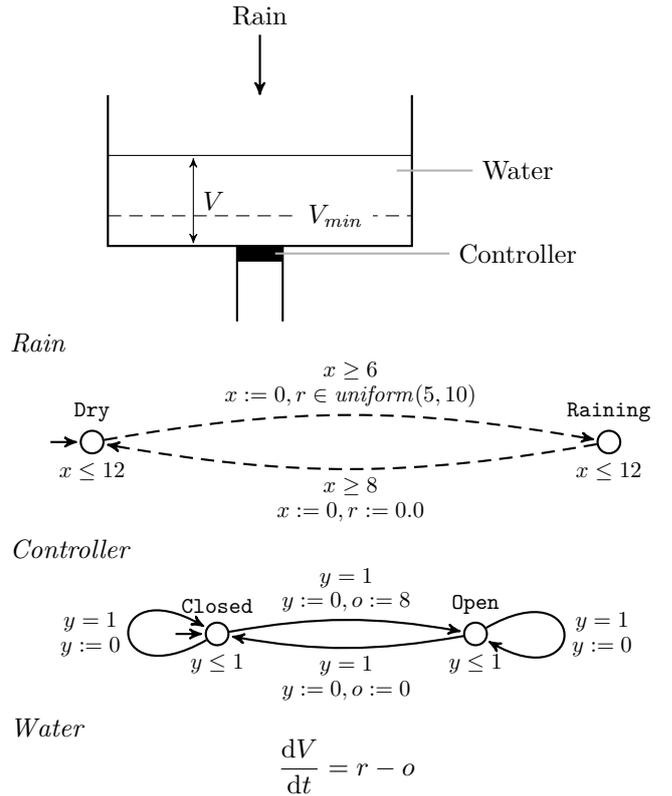

\hll{Figure~\ref{fig:example} shows a small HMDP example. Rain can fall into a storage tank, from which water can be drained with a controlled valve. The uncontrollable environment is modeled with the (stochastic) timed automaton \emph{Rain}. The two locations, depicted with circles, represent the two modes \texttt{Dry} and \texttt{Raining}. The clock variable $x$ keeps track of the duration of modes. The model indicates that the dry mode has a duration between 6 and 12 time units, while the raining mode has a duration between 8 and 12 time units, both uniformly distributed. Once an uncontrollable edge is taken from \texttt{Dry} to \texttt{Raining} (indicated by the dashed arrow), the rain intensity $r$ is chosen uniformly between 5 and 10 volume unit per time unit. The initial location is indicated by the small incoming arrow at location \texttt{Dry}. The initial value of clock variables is assumed to be 0 when not depicted.
}

\hll{The timed automaton \emph{Controller} models the controllable valve, which is either \hlr{in} control mode \texttt{Closed} or \texttt{Open}. The solid edges indicate controllable actions. Clock variable $y$ keeps track of the control period duration, where the control period is set to 1 (see guards $y=1$ on the edges). \hlr{When switching to} the \texttt{Closed} mode, the output flow $o$ is set to 0 volume units per time unit, while \hlr{switching to} the \texttt{Open} mode it is set to 8 volume units per time units.
}

\hll{Finally, the \emph{Water} model describes the evolution of the water volume $V$ over time with a simple differential equation: the volume change is the difference between the water inflow $r$ and the water outflow $o$. For this example, the safety objective could be maintaining a minimal water level $V_{\mathit{min}}$, while the optimization objective is to minimize the expected average (accumulated) water level.}

\subsection{\hl{Strategies for HMDP}}
For a given HMDP, a memoryless and possibly non-deterministic \emph{strategy} $\sigma$ determines which of the control modes can be used in the next period. Formally, a strategy is a function $\sigma: \mathbb{C}\rightarrow 2^{C}$ that returns a nonempty set of allowed control modes in a configuration. A strategy is called \emph{deterministic} if exactly one control mode is permitted in each configuration.

The behavior of an HMDP $\mathcal{M}$ under supervision of a strategy $\sigma$, denoted as the stochastic process $\mathcal{M}\!\upharpoonright\!\sigma$, is defined as follows. A run $\pi$ is a \emph{run according to the strategy $\sigma$} if the controller changes mode according to the strategy $\sigma$, i.e., \hll{$\gamma_i\xrightarrow{\tau_{i+1}} (c_i, u_i, \boldsymbol{x}_{i+1})$ with $c_{i+1}\in\sigma((c_i,u_i, \boldsymbol{x}_{i+1} ))$ and $u_{i+1} = u_i$}. 

A strategy $\sigma$ is called \emph{safe} with respect to a set of configurations $S\subseteq \mathbb{C}$ if for any run $\pi$ according to $\sigma$ all configurations encountered are within the safe set $S$. Note that we require $\gamma_i \in S$ for all $i$ and also $\gamma_i'\in S$ whenever $\gamma_i\xrightarrow{\tau}\gamma_i'$ with $\tau\leq P$. \hl{A safe strategy is called \emph{maximally permissive} if for each configuration it returns the largest set of possible actions~\citep{david_uppaal_2015}.}

The optimality of a strategy can be evaluated for the stochastic process $\mathcal{M}\! \upharpoonright\! \sigma$ with a given optimization variable. Let $H\in\mathbb{R}_{\geq 0}$ be a given time-horizon and $D$ a random variable on finite runs, then $\mathbb{E}_{\sigma,H}^{\mathcal{M},\gamma}(D) \in \mathbb{R}_{\geq 0}$ is the expected value of $D$ on the space of runs of $\mathcal{M} \upharpoonright \sigma$ of length $H$ starting in configuration $\gamma$. For example, $D$ can be the integrated error (or deviation) of a continuous variable with respect to its desired target value.

The goal is to synthesize a safe and optimal strategy $\sigma_{\!\mathit{opt}}$ for a given HMDP $\mathcal{M}$, initial configuration $\gamma$, safety set $S$, optimization variable $D$, and time-horizon $H$. To obtain $\sigma_{\!\mathit{opt}}$, \hll{the tool \stratego} first a \hl{maximally permissive} safe strategy $\sigma_{\!\mathit{safe}}$ is synthesized with respect to $S$. Subsequently, $\sigma_{\!\mathit{opt}}$ is a sub-strategy of $\sigma_{\!\mathit{safe}}$ \hl{(i.e., $\forall \gamma \in\mathbb{C}: \sigma_{\!\mathit{opt}}(\gamma) \subseteq \sigma_{\!\mathit{safe}}(\gamma)$)} that optimizes (minimizes or maximizes) $\mathbb{E}_{\sigma_{\!\mathit{safe}},H}^{\mathcal{M},\gamma}(D)$. \hl{For additive random variables, the optimal sub-strategy of the maximally permissive strategy is deterministic.}

\subsection{\stratego}
We use the modeling tool \stratego~\citep{david_uppaal_2015} for control synthesis. It integrates \uppaal with the two branches \uppaalsmc~\citep{Bulychev_uppaal-smc_2012} (statistical model checking \hll{for stochastic hybrid systems}) and \uppaaltiga\citep{behrmann_uppaal-tiga_2007} (synthesis for timed games). \hlr{Therefore, \uppaal is able to synthesize safe \emph{and} optimal strategies.} To synthesize a safe and optimal strategy $\sigma_{\!\mathit{opt}}$, \stratego first abstracts the HMDP $\mathcal{M}$ into a 2-player timed game, ignoring all stochasticity. A safe strategy $\sigma_{\!\mathit{safe}}$ is afterwards synthesized for this timed game. A simplified version of timed computational tree logic (TCTL)~\citep{behrmann_uppaal-tiga_2007} is used to formulate the safety specification. Subsequently, reinforcement learning is used to obtain an optimal sub-strategy $\sigma_{\!\mathit{opt}}$ based on $\mathcal{M} \upharpoonright \sigma_{\!\mathit{safe}}$ and the given random optimization variable~\citep{david_uppaal_2015}. 

\hl{Several learning algorithms are at the modelers disposal in \stratego. Recently, in~\cite{jaeger_teaching_2019} Q-learning and M-learning were introduced. With Q-learning, sample runs are drawn from the HMDP model and are used afterwards to calculate the so-called Q-values. These values are refined into a new strategy and the previous step is repeated with this new strategy until some termination criteria is met. M-learning works similar to Q-learning, except that the HMDP model is now used to approximate the transition and cost functions, which are used to calculate the Q-values instead of sample runs.} \hll{To efficiently cope with continuous state spaces, \stratego uses online partition refinement techniques.}

\section{Modeling}\label{sect:modeling}

Our model \hll{of a detention pond} consists of the components \hlr{Pond, Controller, Rain, and UrbanCatchment}\footnote{The model can be downloaded from \url{http://doi.org/10.5281/zenodo.4719450}.}. 

\subsubsection{Pond}
The dynamics of the water level $w$ in the pond can be derived using the mass balance equation, see~\cite{thomsen_simplified_2019}. Assuming constant density of water, this translates into a volume balance equation. Using Figure~\ref{fig:systemoverview}, we see that the difference in inflow and outflow contribute to the change in water inside the pond. Therefore,
\begin{equation}\label{eq:vbe}
	Q_{\mathit{in}} - Q_{\mathit{out}} = \frac{\mathrm{d}V}{\mathrm{d}t},
\end{equation}
where $Q_{\mathit{in}}$ is the water inlet flow from the urban drainage system, $Q_{\mathit{out}}$ is the water outlet flow into a nearby stream, and $V$ the water volume of the pond above the permanent water level. \hll{The outlet flow is assumed to more or less constant and equal to the discharge permission, but in reality, there will be non-linear relationship to the water level. This is however simplified in this model.}

The change in volume over time can also be expressed using the geometry of the pond:
\begin{equation}\label{eq:vc}
	\frac{\mathrm{d}V}{\mathrm{d}t} = \frac{\mathrm{d}(wA_p(w))}{\mathrm{d}t},
\end{equation}
where $A_p(w)$ is the pond surface area at height $w$. Equations~(\ref{eq:vbe}) and (\ref{eq:vc}) together describe the dynamics of the pond's water level under `normal' circumstances.

There are two boundary cases that need to be taken into account. The first case is when the outflow is larger than the inflow and the water level reaches the permanent water level. The second case is when the inflow is larger than the outflow and the water level reaches the maximum height $W$ of the pond, which results in an emergency overflow. In both cases, the water level $w$ remains stationary.

Now, Equation~(\ref{eq:vbe}) can be reformulated taking these boundary cases into account:
\begin{equation}
	\frac{\mathrm{d}V}{\mathrm{d}t} = 
	\begin{cases}
		0 & \mbox{if } Q_{\mathit{out}} \geq Q_{\mathit{in}} \wedge w = 0, \\
		0 & \mbox{if } Q_{\mathit{in}} \geq Q_{\mathit{out}} \wedge w = W, \\
		Q_{\mathit{in}} - Q_{\mathit{out}} & \mbox{otherwise.}
	\end{cases}
\end{equation}

\subsubsection{Controller}
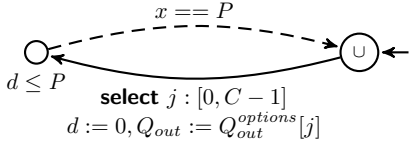
\begin{figure}
	\centering
	\begin{tikzpicture}[->,>=stealth',shorten >=1pt,opacity=1, auto,node distance=5cm,
		thick,main node/.style={circle,draw,font=\sffamily\Large\bfseries},scale=0.85,transform shape]
		
		\node[main node, label={below:$d \leq \mathit{P}$}, minimum size=10pt] (2)  {};
		\node[initial,  initial text ={}, initial where=right,main node, minimum size=10pt] (3) [right of=2] {\footnotesize$\cup$};
		
		
		\path[every node/.style={font=\sffamily}, bend left=\a, align=center]
		(2) edge [dash pattern=on5pt off3pt] node [above] {$x==P$} 	(3)
		(3) edge [] node [below] {\textbf{select} $j:[0,C-1]$\\ $d := 0, Q_{out}:=Q_{out}^{\mathit{options}}[j]$} 	(2);		
	\end{tikzpicture}
	\caption{The model of the controller. The \textbf{select} statement is a simplification of having a set of edges with one for each value in the select range.}\label{fig:controller}
\end{figure}

The controller is able to change the size of the pond outlet valve periodically. Figure~\ref{fig:controller} shows the model of the controller. 

It starts in the urgent location on the right from where it choses with a controllable action one of the $C$ control options. The actual output $Q_{\mathit{out}}$ is set to one of the predefined constant outputs for each mode $Q_{\mathit{out}}^{\mathit{options}}$ and the clock $d$, measuring the duration of the current control period, is reset to $0$. Note that $d$ is local to the controller and does not interfere with clock $d$ from the Rain model.

In the left location, the controller waits until the period with duration $P$ is over, indicated with invariant $d \leq P$. When the controller has waited for $P$ time units, it goes to the right urgent location and above process is repeated for the next control period.

\subsubsection{Rain}
\begin{figure}
	\centering
	\begin{tikzpicture}[->,>=stealth',shorten >=1pt,opacity=1, auto,node distance=8cm,
		thick,main node/.style={circle,draw,font=\sffamily\Large\bfseries},scale=0.85,transform shape]
		\node[initial,  initial text ={}, initial where=left, main node, label={above:$\texttt{Dry}$}, label={below:$d \leq \mathit{dryU}[i]$}, minimum size=10pt] (1) {};
		
		\node[main node,  label={above:$\texttt{Raining}$}, label={below:$d \leq \mathit{rainU}[i]$}, minimum size=10pt] (2) [right of=1] {};
		
		\path[every node/.style={font=\sffamily}, dash pattern=on5pt off3pt, bend left=\a, align=center]
		(1) edge [bend left] node [above] {$d\geq \mathit{dryL}[i]$\\ $d :=0, \mathit{rain} :=\mathit{rain}[i]\cdot\mathit{uniform}(1-\varepsilon, 1+\varepsilon)$} 	(2)
		(2) edge [bend left] node [below] {$d\geq \mathit{rainL}[i]$ \\  $d := 0, i\!+\!+, \mathit{rain} := 0.0$} (1);			
	\end{tikzpicture}
	\caption{The model generating rainfall.}\label{fig:rain}
\end{figure}
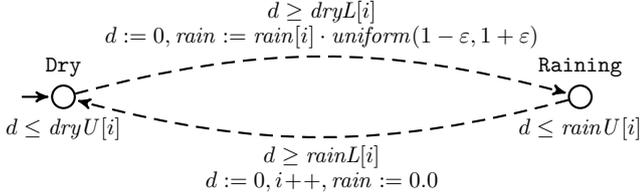

Figure~\ref{fig:rain} shows the rain model \hl{including its uncertainty}. \hlr{It generates the uncontrollable input to the system.} The rain profile is modeled as alternating dry and raining intervals, \hll{each modeled with a location}. For each interval period, indicated with $i$, the duration of the dry (raining) period is bounded between $\mathit{dryL}[i]$ ($\mathit{rainL}[i]$) and $\mathit{dryU}[i]$ ($\mathit{rainU}[i]$), \hl{all being positive integers}. Clock $d$ tracks the duration of the current interval. The actual dry or rain duration is chosen uniformly at random between $\mathit{dryL}[i]$  and $\mathit{dryU}[i]$ or $\mathit{rainL}[i]$ and $\mathit{rainU}[i]$, respectively. 

When it is raining in the $i$-th interval, the actual rain intensity $\mathit{rain}\in\mathbb{R}$ used as input for the UrbanCatchment model is chosen uniformly random between $\mathit{rain}[i]\cdot (1-\varepsilon)$ and $\mathit{rain}[i]\cdot (1+\varepsilon)$, where $\varepsilon$ is a fixed uncertainty factor. \hll{Within an interval, the rain intensity is constant.}

The concrete values for $\mathit{dryL}[i], \mathit{dryU}[i], \mathit{rainL}[i],\allowbreak \mathit{rainU}[i]$, and $\mathit{rain}[i]$ are derived from historical rain data from the~\cite{dmi}. 

\subsubsection{UrbanCatchment}
We model the urban catchment area as a one-layer linear reservoir model (the surface storage of the simplified \hl{`Nedb\o{}r Affstr\o{}mnings Model'} (NAM)), see~\cite{nielsen_numerical_1973}. It is assumed that both the $\mathit{rain}$ and the stored rain water $\hl{S\in\mathbb{R}}$ are uniformly distributed along the urban area, so both become a height measure instead of volume. The time-dependent dynamics of $S$ is given by
\begin{equation}
	\frac{\mathrm{d}S}{\mathrm{d}t} = \mathit{rain} - kS,
\end{equation}
where $k$ is the urban surface reaction factor. \hll{This expression simply states that the change in stored water $S$ depends on the different between the rain falling into the urban area and the storm water leaving it.} The flow (\hll{expressed as a volume per time unit}) from the urban catchment to the pond $Q_{\mathit{in}}$ is given by
\begin{equation}
	Q_{\mathit{in}} = kSA_{uc}
\end{equation} 
with $A_{uc}$ being the urban catchment surface area.

\section{Controller synthesis}\label{sect:synthesis}
\subsection{Problem definition}
The main objective of the controller is to ensure a safe operation of the storm water detention pond. In this context, safety is defined as preventing the pond from overflowing. In case of an overflow event, the water discharge in the nearby stream or river is temporarily much higher than normal. This excessive discharge might have environmental impacts or cause downstream flooding.

We measure overflow with a continuous variable $o$ that represents the accumulated overflow duration. Formally, 
\begin{equation}
	\frac{\mathrm{d}o}{\mathrm{d}t} = 
	\begin{cases}
		1 & \mbox{if } w = W, \\
		0 & \mbox{if } w < W.
	\end{cases}
\end{equation}

The secondary objective is to capture as much urban area particles as possible from the storm water. This is done to prevent contamination of the nearby stream or river. Particles are captured through particle sedimentation onto the pond's floor surface. Hence, the more water in the pond, the more time particles have to be deposited on the floor.

In the model, we associate a cost $c$ to the ability of particle sedimentation. A linear cost function is used such that higher water levels, related to higher possibilities for particle sedimentation, result in lower cost. Formally,
\begin{equation}
	\frac{\mathrm{d}c}{\mathrm{d}t} = 1 - \frac{w}{W}.
\end{equation}
Therefore, $c$ represents the accumulated cost, where a cost of 1 per time unit is associated with $w$ being the permanent water level and a cost of 0 with $w$ being at the maximum height $W$.

Now, the controller synthesis problem is formulated as follows. \hl{Synthesize the safe strategy $\sigma_{\!\mathit{safe}}$ with respect to the safe state set $S$ specified by TCTL predicate $\phi\equiv\square (o = 0)$\footnote{In \stratego, we actually implemented it with $\lozenge (o = 0 \wedge t = H)$, which is equivalent as $o$ is a monotonically increasing variable, and integrated it with Equation~\ref{eq:optimal}.}, i.e., no overflow is encountered in all states of the runs according to the safe strategy. Subsequently, the optimal strategy is calculated with} 
\begin{equation}\label{eq:optimal}
	\sigma_{\!\mathit{opt}} = \hll{\argmin_{\sigma\subseteq \sigma_{\!\mathit{safe}}}}\  \mathbb{E}_{\sigma_{\!\mathit{safe}} ,H}^{\mathcal{M}, \gamma}(c),
\end{equation} 
\hl{where the particle sedimentation cost is minimized while adhering to the synthesized safe strategy $\sigma_{\!\mathit{safe}}$. \stratego is used to synthesize controllers for this problem, where the continuous variables $o$ and $c$ are implemented in a separate component and Equation~(\ref{eq:optimal}) is the optimization query.}

\subsection{Experimental results}

We calibrated our model to the Vilhelmsborg Skov pond south of Aarhus, Denmark. It has an urban catchment area of $A_{\mathit{uc}}= 0.59$ ha, a permitted discharge of $95$ L/s, and an average pond area $A_p = 5,572$ m$^2$ (data from~\cite{thomsen_simplified_2019}). \hl{We estimated \hl{the urban surface reaction factor to $k=0.25$ and} the maximum water level to $W=300$ cm.}

\begin{table}
	\renewcommand{\arraystretch}{1.3}
	\caption{The rain forecast data}
	\label{tab:rain_data}
	\centering
	{\tabulinesep=1.2mm
		\begin{tabu}{|X[0.5,c,m]|X[1,c,m]|X[1,c,m]|X[1,c,m]|X[1,c,m]|X[1.4,c,m]|}
			\tabucline-
			$I$ & $dryL$ [min] & $dryU$ [min] & $rainL$ [min] & $rainU$ [min] & $\mathit{rain}$ [mm/min] \\ \tabucline[1.5pt]-
			1 & 210 & 256 & 27 & 33 & 0.01333 \\ \tabucline-
			2 & 64 & 78 & 21 & 25 & 0.03478 \\ \tabucline-
			3 & 1376 & 1682 & 49 & 61 & 0.02545 \\ \tabucline-
			4 & 168 & 206 & 23 & 29 & 0.02308 \\ \tabucline-
			5 & 203 & 249 & 208 & 254 & 0.00952 \\ \tabucline-
	\end{tabu}}
\end{table}

Historical rain data for the period September 5 - September 7 2019 are used, obtained from the~\cite{dmi}. For each rain event, we averaged the rain intensity. Subsequently, an uncertainty of $\varepsilon=10\%$ is added to the observed interval durations and rain intensities to obtain a weather forecast. In this period, five rain events occurred with varying lengths and intensities. 
Table~\ref{tab:rain_data} shows the obtained rain data used in the model, where September 5 starts dry (so the first rain is expected to start falling between 3.30 am and 4.16 am).

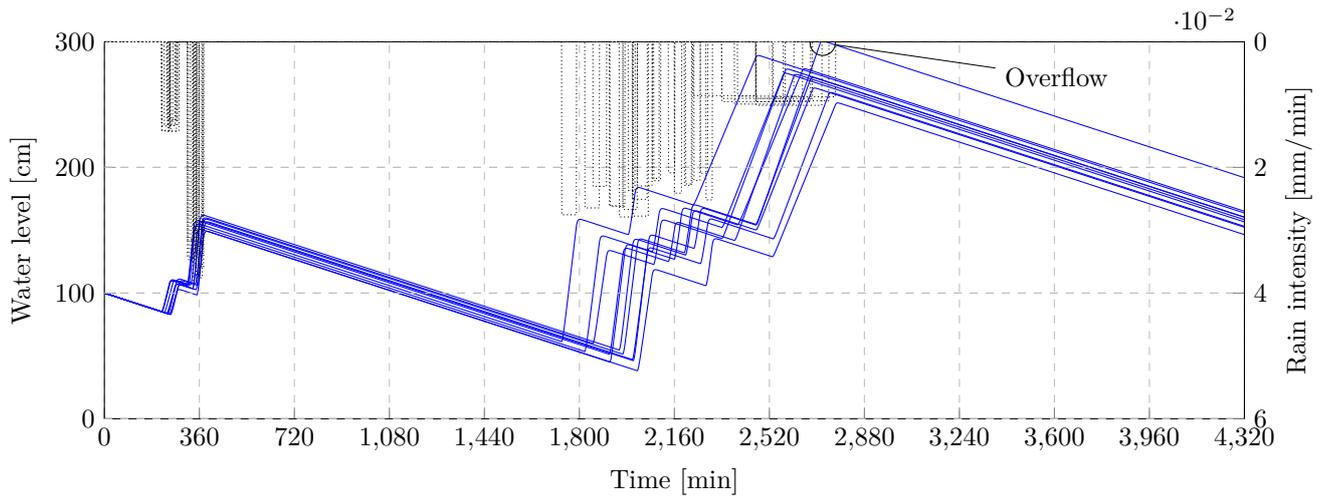
\begin{figure*}
	\centering
	\begin{tikzpicture}
		pgfplotsset{set layers}
		\begin{axis}[
			scale only axis,
			xlabel={Time [min]},
			ylabel={Water level [cm]},
			xmin=0, xmax=4320,
			ymin=0, ymax=300,
			axis y line*=left,  
			axis x line*=bottom, 
			width=15cm,
			height=5cm,
			xtick distance=360,
			xmajorgrids=true,
			ymajorgrids=true,
			grid style=dashed,
			yticklabel style={
				/pgf/number format/precision=0,
				/pgf/number format/fixed,
				/pgf/number format/fixed zerofill,
			},
			]
			
			\foreach \i in {1,2, ...,10} {
				\addplot[
				color=blue,
				]
				table[]{data/sim_static_100_w_\i.txt};
			}	
			
			\node [draw, circle,pin={[pin distance=60, pin edge={black},] 355:{Overflow}}] at (2722,299) {};	
		\end{axis}
		
		\begin{axis}[
			scale only axis,
			ylabel={Rain intensity [mm/min]},
			xmin=0, xmax=4320,
			ymin=0, ymax=0.06,
			axis y line*=right,
			axis x line=none,
			y dir=reverse,
			width=15cm,
			height=5cm,
			xtick distance=360,
			xmajorgrids=true,
			ymajorgrids=true,
			grid style=dashed,
			yticklabel style={
				/pgf/number format/precision=0,
				/pgf/number format/fixed,
				/pgf/number format/fixed zerofill,
			},
			]
			
			\foreach \i in {1,2,...,10} {
				\addplot[
				color=black,
				densely dotted,
				]
				table[]{data/sim_static_100_rain_\i.txt};
			}
		\end{axis}	
	\end{tikzpicture}
	\caption{Ten simulations of the model with the current static control. Blue and solid lines indicate the water level in the pond and black and dotted lines the rain fall. \hl{For one run, overflow occurs around $t=2700$ min, as the maximum water level is $W=300$ cm.} As a quick reference, 1440 minutes is 1 day, and the total time scale is 3 days.}\label{fig:results_static_100}
\end{figure*}

Figure~\ref{fig:results_static_100} shows the results of ten simulated runs \hl{in \stratego} with an initial water height $w=100$ cm and the current static control strategy, i.e., the number of control modes $C$ for the valve is 1. In blue and solid lines the water level is plotted, in \hlr{black and dotted} lines the rain. \hl{The discretization step for the simulations is set to 0.5 minutes.} We observe that one of the ten runs eventually results in an emergency overflow of the pond around 2700 minutes (9 pm on September 6). This is also confirmed by analyzing the expected value of $o$: $\mathbb{E}_{\sigma_{\!\mathit{static}},3 \mbox{ days}}^{\mathcal{M}, w=100}(o) = 1.8 \pm 0.4$, i.e., the pond is expected to be overflowing for 1.8 minutes. 

\begin{figure*}
	\centering
	\begin{tikzpicture}
		pgfplotsset{set layers}
		\begin{axis}[
			scale only axis,
			xlabel={Time [min]},
			ylabel={Water level [cm]},
			xmin=0, xmax=4320,
			ymin=0, ymax=300,
			axis y line*=left,  
			width=15cm,
			height=5cm,
			xtick distance=360,
			xmajorgrids=true,
			ymajorgrids=true,
			grid style=dashed,
			yticklabel style={
				/pgf/number format/precision=0,
				/pgf/number format/fixed,
				/pgf/number format/fixed zerofill,
			},
			]
			
			\foreach \i in {1,2, ...,10} {
				\addplot[
				color=blue,
				]
				table[]{data/sim_dynamic_100_w_\i.txt};
			}		
		\end{axis}
	\end{tikzpicture}
	\caption{Ten simulations of the model with optimal dynamic control. As a quick reference, 1440 minutes is 1 day, and the total time scale is 3 days.}\label{fig:results_dynamic_100}
\end{figure*}
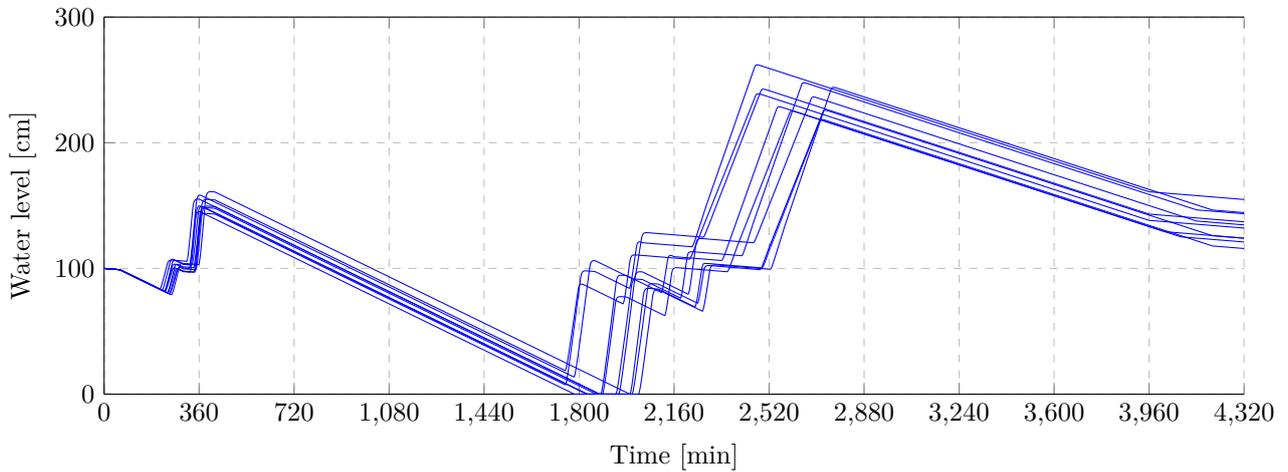

An actively controlled valve can have three different modes: small, medium, and large. We set the medium setting to the current static output flow capacity of $95$ L/s. The low setting is $0.25$ times medium and high $1.5$ times medium. Due to power constraints, the valve can only change once every hour, so $P=60$ min. \hl{We use Q-learning to synthesize strategies with the learning parameters set to 40 successful runs, a maximum of 100, 20 good runs, and 20 runs to evaluate (the first four learning parameters in \stratego).}

With these control modes, we synthesized an optimal controller using Equation~(\ref{eq:optimal}). Figure~\hlr{\ref{fig:results_dynamic_100}} shows ten runs \hl{in \stratego} of the model using the synthesized optimal controller. As can be seen from the figure, in order to ensure safety, the controller keeps the water level in the pond lower than the static controller. For some runs the pond water level even reaches the permanent water level, i.e., it cannot go lower. 

Yet, having a safe strategy comes with higher cost for particle sedimentation. For static control, the expected cost is $\mathbb{E}_{\sigma_{\!\mathit{static}},3 \mbox{ days}}^{\mathcal{M}, w=100}(c) = 2026 \pm 8$, while for the optimal control it is $\mathbb{E}_{\sigma_{\!\mathit{opt}}3 \mbox{ days}}^{\mathcal{M}, w=100}(c) = 2480 \pm 8$. This is an increase of 22\%, \hll{but accepted as our primary aim is to avoid flooding}.


\begin{figure*}
	\centering
	\begin{tikzpicture}
		pgfplotsset{set layers}
		\begin{axis}[
			scale only axis,
			xlabel={Time [min]},
			ylabel={Water level [cm]},
			xmin=0, xmax=4320,
			ymin=0, ymax=300,
			axis y line*=left,  
			width=15cm,
			height=5cm,
			xtick distance=360,
			xmajorgrids=true,
			ymajorgrids=true,
			grid style=dashed,
			yticklabel style={
				/pgf/number format/precision=0,
				/pgf/number format/fixed,
				/pgf/number format/fixed zerofill,
			},
			]
			
			\foreach \i in {1,2, ...,10} {
				\addplot[
				color=blue,
				dashed,
				]
				table[]{data/sim_static_0_w_\i.txt};
				
				\addplot[
				color=red,
				]
				table[]{data/sim_dynamic_0_w_\i.txt};
			}		
		\end{axis}
	\end{tikzpicture}
	\caption{Results for initial water level $w=0$ cm. In blue and dashed lines are ten simulations of the model with static control and in red and solid lines ten with optimal dynamic control. As a quick reference, 1440 minutes is 1 day, and the total time scale is 3 days.}\label{fig:results_dynamic_0}
\end{figure*}
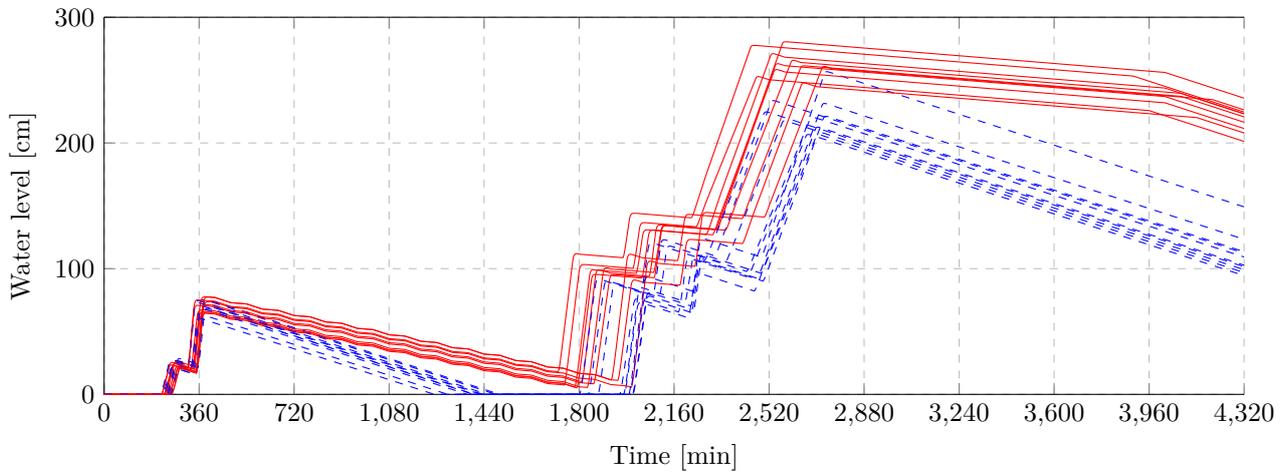

Figure~\ref{fig:results_dynamic_0} shows simulation results for the case that the initial water level is set to 0 cm. We notice that the synthesized optimal strategy now lowers the output valve setting compared to the previous experiment. It is interesting to see that between approximately 360 min and 1800 min, the optimal strategy is switching between the low and medium settings: it tries to create an output flow that is between those settings, so it reaches approximately $w=0$ when the next rain is expected to start. \hll{This indicates that having a fourth output valve setting might be beneficial with respect to saving energy of switching between valve settings.} The cost for particle sedimentation indicates an improvement of 18\% compared to static control: $\mathbb{E}_{\sigma_{\!\mathit{static}},3 \mbox{ days}}^{\mathcal{M}, w=0}(c) = 2945 \pm 6$ for static control and $\mathbb{E}_{\sigma_{\!\mathit{opt}},3 \mbox{ days}}^{\mathcal{M}, w=0}(c) = 2409 \pm 9$ for optimal dynamic control.

Finally, \stratego can also report when no optimal and safe strategy can be synthesized. For example, if the initial water level is $w = 150$ cm, the query for Equation~(\ref{eq:optimal}) cannot be satisfied. This means that no safe strategy can be found against the uncontrollable opponent (in this case the weather forecast) such that the pond will never overflow. This can be useful in predicting when emergency overflows will occur, acting as a warning system for the pond's operational organization, \hl{such that they can take additional measures at the pond in question, \hll{like maximizing the discharge flow,} or in the potentially affected area downstream}.

\section{Conclusion and future challenges}\label{sect:conclusion}
We applied formal controller synthesis to automatically derive controllers for storm water detention ponds where the water discharge into the nearby stream can be regulated. We showed that the problem can be modeled as a hybrid Markov decision process, such that symbolic and reinforced learning techniques from \stratego can be applied. Simulation results of an existing detention pond in Denmark shows that safe and near-optimal active controllers can be synthesized. 

This first step opens several future research directions. First, weather forecasts change over time \hll{and are increasingly used in urban hydrology research~\citep{thorndahl_weather_2017}}. Therefore, the presented model setup should be adapted to an on-line model-predictive control setting. Second, to increase the explainability of the synthesized strategies, it is to be investigated whether exporting strategies to decision trees, see~\cite{ashok_sos_2019}, is possible. \hlr{Third, it would be interesting to validate the approach with real-life data. Yet, actual data of water levels in ponds are scarce.} Finally, only a single storm water detention pond is analyzed in isolation from the discharge stream. It is interesting to see whether collaborative strategies can be synthesized for a collection of detention ponds all discharging into the same stream.

{\small
\bibliography{ref}             
}
\end{document}